\documentclass[twocolumn,aps,prl,superscriptaddress]{revtex4}

\usepackage{graphicx}
\usepackage{dcolumn}

\usepackage{xspace}
\usepackage{relsize}
\def\babar{\mbox{\slshape B\kern-0.1em{\smaller A}\kern-0.1em
    B\kern-0.1em{\smaller A\kern-0.2em R}}\xspace}

\newcommand{\tev}{\ensuremath{\mathrm{\,Te\kern -0.1em V}}\xspace}
\newcommand{\gev}{\ensuremath{\mathrm{\,Ge\kern -0.1em V}}\xspace}
\newcommand{\mev}{\ensuremath{\mathrm{\,Me\kern -0.1em V}}\xspace}
\newcommand{\kev}{\ensuremath{\mathrm{\,ke\kern -0.1em V}}\xspace}
\newcommand{\ev}{\ensuremath{\mathrm{\,e\kern -0.1em V}}\xspace}
\newcommand{\gevc}{\ensuremath{{\mathrm{\,Ge\kern -0.1em V\!/}c}}\xspace}
\newcommand{\mevc}{\ensuremath{{\mathrm{\,Me\kern -0.1em V\!/}c}}\xspace}
\newcommand{\gevcc}{\ensuremath{{\mathrm{\,Ge\kern -0.1em V\!/}c^2}}\xspace}
\newcommand{\mevcc}{\ensuremath{{\mathrm{\,Me\kern -0.1em V\!/}c^2}}\xspace}

\newcommand{\lll}     {\ensuremath{\ell^{-}\ell^{+}\ell^{-}}\xspace}
\newcommand{\eee}     {\ensuremath{e^-\!e^+\!e^-}\xspace}
\newcommand{\eemw}    {\ensuremath{\mu^+\!e^-\!e^-}\xspace}
\newcommand{\eemr}    {\ensuremath{\mu^-\!e^+\!e^-}\xspace}
\newcommand{\emmw}    {\ensuremath{e^+\!\mu^-\!\mu^-}\xspace}
\newcommand{\emmr}    {\ensuremath{e^-\!\mu^+\!\mu^-}\xspace}
\newcommand{\mmm}     {\ensuremath{\mu^-\!\mu^+\!\mu^-}\xspace}

\newcommand{\taulll}  {\ensuremath{\tau^{-}\!\to\lll}\xspace}
\newcommand{\taullls}  {\ensuremath{\tau\!\to\ell\ell\ell}\xspace}
\newcommand{\dEdM}    {\ensuremath{(\Delta M, \Delta E)}\xspace}
\newcommand{\epem}    {\ensuremath{e^+e^-}\xspace}
\newcommand{\mpmm}    {\ensuremath{\mu^+\mu^-}\xspace}
\newcommand{\tptm}    {\ensuremath{\tau^+\tau^-}\xspace}
\newcommand{\qqbar}   {\ensuremath{q\overline q}\xspace}

\def\invfb            {\ensuremath{\mbox{\,fb}^{-1}}\xspace}
\def\pt               {\mbox{$p_T$}\xspace}
\def\kk2f             {\mbox{\tt KK2f}\xspace}
\def\tauola           {\mbox{\tt Tauola}\xspace}

\newcommand{\Nobs}    {\ensuremath{N_{\rm obs}}\xspace}
\newcommand{\Nbgd}    {\ensuremath{N_{\rm bgd}}\xspace}
\newcommand{\Nul}     {\ensuremath{N_{\rm UL}^{90}}\xspace}
\newcommand{\BRul}    {\ensuremath{\BR_{\rm UL}^{90}}\xspace}
\newcommand{\tensev}  {\ensuremath{\times 10^{-7}}\xspace}
\newcommand{\BR}      {\ensuremath{{\cal B}}}
\newcommand{\Bfivepr} {\ensuremath{\BR(\tau^-\to 3h^-2h^+ \nu_\tau)}\xspace}

\newcommand{\jprlBase}       {Phys.\ Rev.\ Lett.\xspace}
\newcommand{\jprl}      [1]  {\jprlBase\ {\bf #1}}
\newcommand{\nimBaseA}       {Nucl.\ Instr.\ Meth.\xspace}
\newcommand{\nima}      [1]  {\nimBaseA~A~{\bf #1}}
\newcommand{\npBase}         {Nucl.\ Phys.\xspace}
\newcommand{\npb}       [1]  {\npBase\ B~{\bf #1}}
\newcommand{\epjBase}        {Eur.\ Phys.\ Jour.\xspace}
\newcommand{\epjc}      [1]  {\epjBase\ C~{\bf #1}}
\newcommand{\npbps}     [1]  {{Nucl.\ Phys.\ B~Proc.\ Suppl.\ {\bf #1}}}

\newcommand{\BABARPubYear}     {04}
\newcommand{\BABARPubNumber}  {122}
\newcommand{\SLACPubNumber} {10831}

\begin{document}

\preprint{\babar-PROC-\BABARPubYear/\BABARPubNumber} 
\preprint{SLAC-PUB-\SLACPubNumber} 

%% Needed in final document
\begin{flushleft}
\babar-PUB-\BABARPubYear/\BABARPubNumber\\
SLAC-PUB-\SLACPubNumber\\
%hep-ex/\LANLNumber\\[10mm]
\end{flushleft}

\title{Results on Tau Physics from \babar}

\author{Eric Torrence}
\affiliation{Physics Department, 1274 University of Oregon, Eugene OR 97403, USA
\\E-mail: {\tt torrence@physics.uoregon.edu}}
\collaboration{Representing the \babar\ Collaboration}
\noaffiliation

\begin{abstract}
Recent results on tau physics from \babar are reviewed.
Limits on lepton-flavor violation in the tau decay process
\taulll are presented based on 91.6\invfb of data.
In all six decay modes considered, the numbers of events found 
in data are compatible with the background expectations, and
upper limits on the branching fractions are set in the range 
$(1-3)\tensev$ at 90\% CL.
A preliminary measurement of the five prong branching fraction
based on 110.7\invfb of data is presented with the
result $\Bfivepr = (8.52\pm0.09\pm0.40)\times 10^{-4}$.
A preliminary measurement of the tau lifetime based on 30\invfb
of data is presented where a lifetime of $290.8\pm1.5\pm1.6$~fs is measured.
\end{abstract}

\maketitle

\section{Introduction}%1

The tau lepton is in many ways a laboratory in its own right.
As the heaviest third generation lepton, it is a natural place
to look for new physics arising at higher mass scales.
It is also the only lepton which can decay to hadrons providing 
a unique environment to test QCD in both the first and second 
generations.
This paper will review some recent \babar results including 
a published search for Lepton-flavor violation (LFV) in the 
decay \taullls \cite{babarlll}, a preliminary measurement of the 
five-prong branching fraction \cite{babar5pr},
and a preliminary result on the tau lifetime.

The PEP-II storage ring, running at the $\Upsilon(4S)$ resonance,
has performed very well and \babar has recorded 244\invfb
of data through July 2004.
With an expected cross section for tau pairs at the PEP-II collision
energy of $\sigma_{\tau\tau} = (0.89\pm0.02)$ nb \cite{kk},
\babar is also a tau factory, with almost 220 million \tptm
events recorded.
The \babar detector is described in detail elsewhere \cite{detector}.

\section{Lepton Flavor Violation}

Lepton-flavor violation involving charged leptons has 
never been observed, and stringent experimental limits 
exist from muon branching fractions:
$\BR(\mu\to e\gamma) < 1.2 \times 10^{-11}$
and $\BR(\mu\to eee) < 1.0 
\times 10^{-12}$ at 90\% confidence level (CL) \cite{brooks99,sindrum88}.
Recent results from neutrino oscillation experiments \cite{neut} 
show that LFV does indeed occur, although the branching fractions 
expected in charged lepton decay due to neutrino mixing alone 
are probably no more than $10^{-14}$ \cite{pham98}.

Many extensions to the Standard Model (SM), particularly models 
seeking to describe neutrino mixing, predict enhanced LFV in tau 
decays over muon decays with branching fractions from 
$10^{-10}$ up to the current experimental limits \cite{ma02}.
Observation of LFV in tau decays would be a 
clear signature of non-SM physics, while improved 
limits will provide further constraints on theoretical models.
Prior to the start of the B-factories, the best limits on LFV
in tau decays came from CLEO: 
$\BR(\tau\to\mu\gamma) < 1.1\times 10^{-6}$ \cite{cleomg}.

In this analysis, all possible lepton combinations consistent 
with charge conservation are considered, leading to six distinct
decay modes (\eee, \eemw, \eemr, \emmw, \emmr, \mmm) \cite{cc}.
The signature of this process is three charged 
particles, each identified as either an electron or muon,
with an invariant mass and energy equal to that of the parent 
tau lepton.
Candidate signal events are required to have
a ``1-3 topology,'' where one tau decay yields three
charged particles (3-prong), while the second tau
decay yields one charged particle (1-prong).

Each of the charged particles found in the 3-prong 
hemisphere must be identified as either an electron
or muon candidate.
The particle identification (PID) requirements alone
are not sufficient to suppress certain backgrounds, 
particularly those from higher-order radiative 
Bhabha and \mpmm\ events that can have four leptons 
in the final state, and additional selection criteria
are applied to reduce these backgrounds.

\begin{figure}
 \resizebox{\columnwidth}{!}{%
\includegraphics{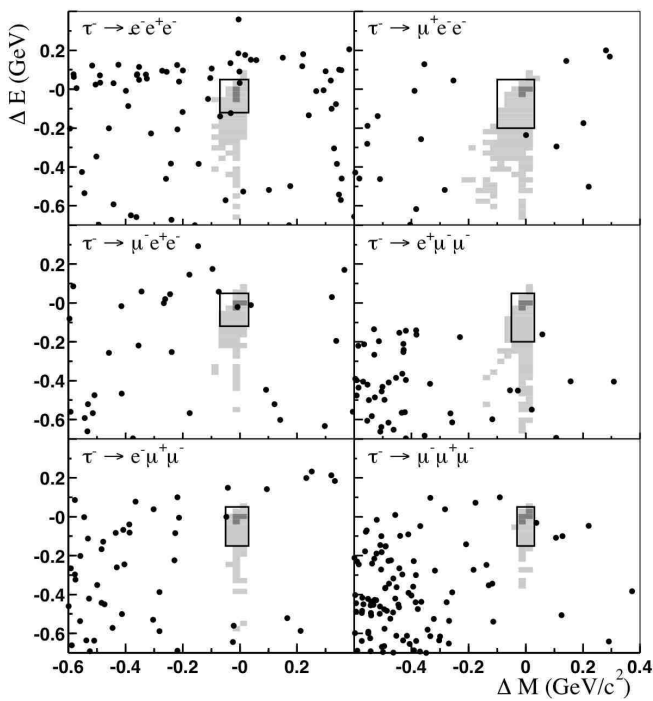}}
\caption{Observed data shown as dots in the \dEdM\ plane and 
the boundaries of the signal region for each decay mode.
The dark and light shading indicates contours containing
50\% and 90\% of the selected MC signal events, respectively.}
\label{fig1}
\end{figure}

Signal events are expected to have an invariant mass and total
energy in the 3-prong hemisphere consistent with the neutrino-less
decay of a tau lepton.
The difference in reconstructed and expected energy $(\Delta E)$ 
and mass $(\Delta M)$ are calculated in the \epem\ rest frame 
from the observed track momenta assuming the corresponding lepton 
masses for each decay mode.
A LFV signal would peak at the origin in the \dEdM\ plane, although
detector resolution and radiative effects broaden the expected 
distribution.
Rectangular signal regions are defined in this \dEdM\ plane for each 
decay mode, and are shown along with observed data and expected
signal distributions in Figure~\ref{fig1}.
To avoid bias, a blinded analysis procedure was adopted
with the number of data events in the signal region
remaining unknown until the selection criteria 
were finalized and all cross checks were performed.

There are three main classes of background remaining after
the selection criteria are applied: low multiplicity \qqbar 
events (mainly continuum light-quark production), QED events 
(Bhabha and \mpmm), and SM \tptm events.
These three background classes have distinctive distributions
in the \dEdM\ plane, and expected background rates are determined
by fitting three PDFs, one for each background type, to the data
observed in a grand sideband region shown as the area of 
Figure~\ref{fig1} excluding the signal region.
The shape of these PDFs are determined from Monte Carlo (MC) samples 
or data control samples.

The primary systematic uncertainties involve the efficiency of the
PID criteria and the uncertainties in determining the background
fractions.
The PID efficiencies are not estimated from MC samples, but rather
are measured directly from data control samples.
The background uncertainty is dominated by the statistical power
of the data in the grand sideband region.

The numbers of events observed (\Nobs) and the 
background expectations (\Nbgd) are shown in Table~\ref{tab:results}
along with the signal efficiency and expected background rates in 
each final state.
No significant excess found in any decay mode, and
upper limits on the branching fractions are calculated according to 
$\BRul = \Nul/(2 \varepsilon {\cal L} \sigma_{\tau\tau})$, where $\Nul$
is the 90\% CL upper limit for the number of signal events when 
\Nobs\ events are observed with \Nbgd\ background events expected.
The branching fraction upper limits have been calculated including
all uncertainties using the technique of 
Cousins and Highland \cite{cousins92} following the implementation of 
Barlow \cite{barlow02}.

\begin{table*}
\begin{center}
\caption{Efficiency estimates, number of expected background events (\Nbgd),
number of observed events (\Nobs), and branching fraction upper limits 
for each decay mode.  
}
\begin{tabular}{|l|cccccc|}
\hline
Mode   & \eee         & \eemw        & \eemr & \emmw        & \emmr        & \mmm \\
\hline
Eff. [\%] &$7.3\pm 0.2$&$11.6\pm 0.4$&$7.7\pm 0.3$ &$9.8\pm 0.5$&$6.8\pm 0.4  $&$6.7\pm 0.5$\\
\hline
\qqbar bg.     &$0.67        $&$0.17        $&$0.39$ &$0.20        $&$0.19        $&$0.29$\\
{\sc QED} bg.        &$0.84        $&$0.20        $&$0.23$ &$0.00        $&$0.19        $&$0.01$ \\
$\tau\tau$ bg.    &$0.00        $&$0.01        $&$0.00$  &$0.01        $&$0.01        $&$0.01$\\
\hline
\Nbgd           &$1.51\pm 0.11$&$0.37\pm 0.08$&$0.62\pm 0.10$ &$0.21\pm 0.07$&$0.39\pm 0.08$&$0.31\pm 0.09$\\
\Nobs           &$1           $&$0           $&$1$ &$0           $&$1           $&$0$\\
\hline
\BRul            &$2.0\tensev  $&$1.1\tensev  $&$2.7\tensev$ &$1.3\tensev  $&$3.3\tensev  $&$1.9\tensev$ \\
\hline
\end{tabular}
\label{tab:results}
\end{center}
\end{table*}

The 90\% CL upper limits on the \taulll\ branching fractions, shown
in Table~\ref{tab:results}, are in the range $(1-3)\times10^{-7}$.
Belle has also produced similar results on a data sample of 
87.1\invfb \cite{bellelll}.
As both limits are statistically limited, these results can be
directly combined taking into account relative luminosities, efficiencies,
and backgrounds in each channels.
The individual limits and combined results are shown in Table~\ref{tab:comb}.
With the first data from the B-factories, these limits have been improved
by an order of magnitude the level of $10^{-7}$.
With over 500\invfb expected from each machine in the next few years,
and little background expected, these limits should approach $10^{-8}$
in the near future.
 
\begin{table}
\begin{center}
\caption{Limits on LFV in \taulll\ from \babar and Belle 
($\times 10^{-7}$ at 90\% CL).}  
\begin{tabular}{|l|ccc|}
\hline
Mode & \babar \cite{babarlll} & Belle \cite{bellelll} & Combined \\
\hline
\eee  & 2.0 & 3.5 & 1.5 \\
\eemw & 1.1 & 2.0 & 0.6 \\
\eemr & 2.7 & 1.9 & 1.2 \\
\emmw & 1.3 & 2.0 & 0.7 \\
\emmr & 3.3 & 2.0 & 1.4 \\
\mmm  & 1.9 & 2.0 & 0.8 \\
\hline
\end{tabular}
\label{tab:comb}
\end{center}
\end{table}

\section{Five-prong Tau Decays}

The semi-leptonic tau decay provides a particularly clean environment
to study QCD and the hadronization process.
With the large data sample available at \babar, precision studies of
hadronic structure can now be carried out in final states which were
previously statistically limited.
The previous best measurement of \Bfivepr performed by CLEO was based
on a sample of 295 signal events selected in 1.7\invfb of data \cite{cleo5pr}.
In this analysis, a sample of nearly 15,869 events has been selected
from a data sample of 110.7\invfb.

\begin{figure}
 \resizebox{\columnwidth}{!}{%
\includegraphics{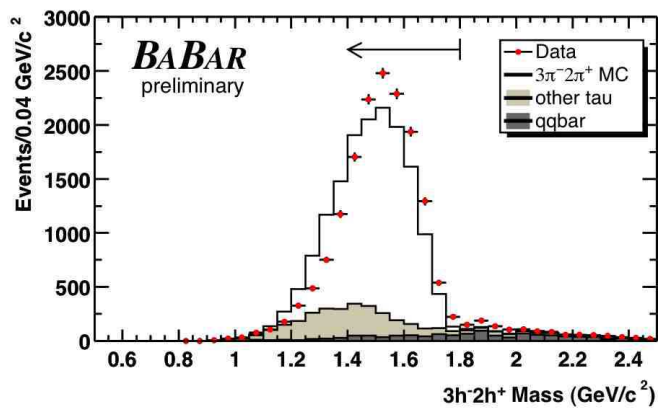}}
\caption{Invariant mass distribution reconstructed in 5-prong tau decays
compared to a phase space MC prediction.}
\label{fig2}
\end{figure}

\begin{figure}
 \resizebox{\columnwidth}{!}{%
\includegraphics{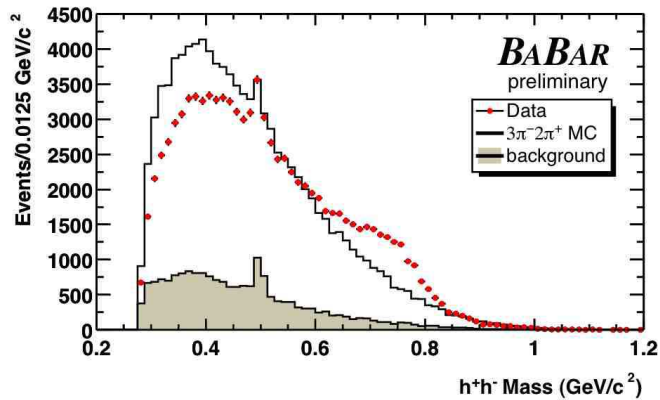}}
\caption{Invariant mass of all opposite-signed pairs compared to a 
phase space MC predicton.}
\label{fig3}
\end{figure}

Events are selected with a 1-5 topology, with an electron or muon
tag on the one-prong hemisphere, additional requirements on the
thrust, \pt, and missing energy in the event.
Electron, $\pi^0$, and conversion candidates are not allowed in
the signal hemisphere.
No attempt has been made to classify hadrons as either pions
or kaons.
This selection achieves an efficiency of 7.5\% with a purity of
79\%.
The majority of the backgrounds come from other tau decays with
either five charged tracks and additional neutral pions, or tau
decays involving $K_s$ mesons.
A preliminary branching fraction measurement of 
$\Bfivepr = (8.52\pm0.09\pm0.40)\times 10^{-4}$
is obtained which represents more than a factor of two improvement
over the previous best measurement by CLEO.

Beyond the branching fraction, the large data sample allows a first
detailed study of the hadronic structure in five-prong tau decays.
Figure~\ref{fig2} shows the invariant mass distribution compared to
a phase-space MC simulation provided by \tauola \cite{tauola}.
Clearly this simple model can not replicate the observed mass 
distribution.
Figure~\ref{fig3} further shows the invariant mass of all 
opposite-signed pairs (six entries per event) again compared 
to the phase space model currently in \tauola.
There is clear evidence in the data of $\rho$ production, which
while not surprising, is the first such observation in five-prong
decays.

\section{Tau Lifetime}

\babar also has performed a measurement of the tau lepton lifetime.
Again using the high statistics available, a very pure selection
of 1-3 events is performed and the $r-\phi$ decay length is 
reconstructed from the primary interaction point and the 3-prong 
vertex.
To reduce systematic uncertainties, the simple mean of the decay 
length distribution is used to extract the lifetime.
A preliminary result based on 30\invfb finds a tau lifetime of
$290.8\pm1.5\pm1.6$ fs, which is competitive with lifetime measurements
performed at LEP.

\section{Conclusions}

First results from \babar on tau physics have been presented with
limits on LFV decays, hadronic structure, and the tau lifetime.
This represents the start of a rich physics program made possible
by the well understood detector and very large data sample available
at \babar.
Additional new preliminary results including further LFV decay modes,
more hadronic structure measurements, and an update of the lifetime
will be made at the Tau04 workshop in September \cite{tau04}.


\begin{thebibliography}{99}

\bibitem{babarlll}
%BaBar collaboration, 
B.~Aubert {\it et al.},
%``Search for lepton flavor violation in the decay tau- $\to$ l- l+ l-,''
\jprl{92}, 121801 (2004).

\bibitem{babar5pr}
%BaBar Collaboration, 
B.~Aubert {\it et al.}, hep-ex/0408050.

\bibitem{kk}
B.~F.~Ward, S.~Jadach, and Z.~Was,
%``Precision calculation for e+ e- $\to$ 2f: The K K MC project,''
Nucl.\ Phys.\ Proc.\ Suppl.\  {\bf 116}, 73 (2003).
%[arXiv:hep-ph/0211132].

\bibitem{detector}
%B.~Aubert {\it et al.}  [BABAR Collaboration],
%%``The BaBar detector,''
%Nucl.\ Instrum.\ Meth.\ A {\bf 479}, 1 (2002)
%[arXiv:hep-ex/0105044].
%\babar\ Collaboration, 
B.~Aubert {\it et al.}, \nima{479}, 1 (2002).

\bibitem{brooks99}
%M.~L.~Brooks {\it et al.} [MEGA/LAMPF Collaboration], 
%% NEW LIMIT FOR THE FAMILY NUMBER NONCONSERVING DECAY MU+ ---> E+ GAMMA.
%Phys.~Rev.~Lett. {\bf 83}, 1521 (1999).
%MEGA/LAMPF Collaboration, 
M.~L.~Brooks {\it et al.}, \jprl{83}, 1521 (1999).

\bibitem{sindrum88}
%U.~Bellgardt {\it et al.} [SINDRUM Collaboration], 
%% SEARCH FOR THE DECAY MU+ ---> E+ E+ E-.
%Nucl.~Phys.~B {\bf 299} 1 (1988).
%SINDRUM Collaboration, 
U.~Bellgardt {\it et al.}, \npb{299}, 1 (1998).

\bibitem{neut}
%K2K Collaboration, 
M.~H.~Ahn {\it et al.}, \jprl{90}, 041801 (2003);
%KamLAND Collaboration, 
K.~Eguchi {\it et al.}, \jprl{90}, 021802 (2003);
%SNO Collaboration, 
Q.~R.~Ahmad {\it et al.}, \jprl{89}, 011301 (2002);
%Super-Kamiokande Collaboration, 
Y.~Fukuda {\it et al.}, \jprl{81}, 1562 (1998).

\bibitem{pham98}
X.~Y.~Pham, \epjc{8}, 513 (1999).

\bibitem{ma02}
E.~Ma,
%``Theoretical expectations for rare and forbidden tau decays,''
\npbps{123}, 125 (2003).

\bibitem{cleomg}
S.~Ahmed {\it et al.},
%``Update of the search for the neutrinoless decay tau $\to$ mu gamma,''
Phys.\ Rev.\ D {\bf 61}, 071101 (2000).

\bibitem{cc}
Throughout this paper, charge conjugate decay modes also are implied.

\bibitem{cousins92}
R.~D.~Cousins and V.~L.~Highland,
%``Incorporating systematic uncertainties into an upper limit,''
Nucl.\ Instrum.\ Meth.\ A {\bf 320}, 331 (1992).

\bibitem{barlow02}
R.~Barlow,
%``A calculator for confidence intervals,''
Comput.\ Phys.\ Commun.\  {\bf 149}, 97 (2002).
%[arXiv:hep-ex/0203002].

\bibitem{bellelll}
Y.~Yusa {\it et al.},
%``Search for neutrinoless decays tau $\to$ 3l,''
Phys.\ Lett.\ B {\bf 589}, 103 (2004).

\bibitem{cleo5pr}
D.~Gibaut {\it et al.},
%``Study of the five charged pion decay of the tau lepton,''
Phys.\ Rev.\ Lett.\  {\bf 73}, 934 (1994).

\bibitem{tauola}
S.~Jadach, Z.~Was, R.~Decker, and J.~H.~Kuhn,
%``The tau decay library TAUOLA: Version 2.4,''
Comput.\ Phys.\ Commun.\  {\bf 76}, 361 (1993).

\bibitem{tau04}
8th International Workshop on Tau-Lepton Physics (Tau04),
14-17 Sep 2004, Nara, Japan.

\end{thebibliography}
\end{document}